\documentclass{optica-article}

\journal{boe}

\usepackage{lineno}
\usepackage{cite}
\usepackage{algorithmic}
\usepackage{graphicx}
\usepackage{textcomp}
\usepackage{hyperref}
\usepackage{multirow}
\usepackage{afterpage}
\usepackage{lipsum}
\usepackage{array}
\usepackage{caption}
\usepackage{subcaption}
\usepackage{threeparttable}
\usepackage{pdfpages}
\usepackage{tabularx}
\usepackage{placeins}
\usepackage{amsmath}
\usepackage{ragged2e}

\begin{document}

\title{Single Frame Laser Diode Photoacoustic Imaging: Denoising and Reconstruction}

\author{Vincent Vousten\authormark{1,2,\textdaggerdbl}, Hamid Moradi\authormark{1,\textdaggerdbl}, Emad M. Boctor\authormark{3}, and Septimiu E. Salcudean\authormark{1,*}}

\address{\authormark{1}{University of British Columbia, Dept. of Electrical and Computer Engineering, Vancouver BC, Canada} \\
\authormark{2}Eindhoven University of Technology, Dept. of Biomedical Engineering, Eindhoven, the Netherlands\\
\authormark{3}John Hopkins University, Whiting School of Engineering, Baltimore, MD, United States\\
\authormark{\textdaggerdbl}Equal contribution}

\email{\authormark{*}\href{mailto:tims@ece.ubc.ca}{tims@ece.ubc.ca}} 


\begin{abstract}
A new development in photoacoustic (PA) imaging has been the use of compact, portable and low-cost laser diodes (LDs), but LD-based PA imaging suffers from low signal intensity recorded by the conventional transducers. A common method to improve signal strength is temporal averaging, which reduces frame rate and increases laser exposure to patients. To tackle this problem, we propose a deep learning method that will denoise the PA images before beamforming with a very few frames, even one. We also present a deep learning method to automatically reconstruct point sources from noisy pre-beamformed data. Finally, we employ a strategy of combined denoising and reconstruction, which can supplement the reconstruction algorithm for very low signal-to-noise ratio inputs.
\end{abstract}

\section{Introduction}\label{sec:introduction}

Photoacoustic imaging (PAI) is a novel and fast-emerging biomedical imaging technique \cite{Attia2019,beard2011biomedical,Jeon2019a,Steinberg2019,Xu2006}. Short pulses of non-ionizing, infrared light are used to excite ultrasound (US) waves in biological tissue through the photoacoustic (PA) effect. These generated US waves are collected using US transducers. This allows for ultrasonic resolution with the contrast induced by light absorption in tissue \cite{Xu2006PhotoacousticBiomedicine}. PAI has shown a great potential in tissue characterization \cite{Yang2011,Gao2014}, needle tracking \cite{Su2010PhotoacousticTissue,LedijuBell2018Photoacoustic-basedTip}, and cancer detection \cite{Xu2006PhotoacousticBiomedicine,Dogra2013MultispectralResults,Ishihara2017,LedijuBell2015TransurethralImaging}.

Compact, portable, inexpensive, and high frequency, but low-power, laser diodes (LDs) have been a new approach for PAI \cite{Kolkman2006InDiode,Upputuri2018FastReview}. LDs would enable clinical applications that are difficult to envisage with the bulky, high-power Nd:YAG lasers that are currently in use. However, those lasers provide a superior signal due to the higher power, and consequently, provide a much better signal-to-noise ratio (SNR) than LD-based PAI.

Here, we explore LD-based PAI with the conventional ultrasound transducers, which are widely used for B-mode imaging, for spot imaging where a single optical absorber is illuminated \cite{Moradi2021ARecovery}. One application for LD-based PAI is for needle tip localization during surgical intervention \cite{Su2010PhotoacousticTissue, LedijuBell2018Photoacoustic-basedTip}. In this application, the needle tip can be designed as an optical source, generating US waves through the PA effect. Tracking the needle tip can also help in biopsies \cite{Wang2019,Kim2018MultimodalGuidance}. This method is also relevant to prostate nerve sensing to preserve the critical structures during prostatectomy. This procedure is commonly performed with robot assistance, which requires real-time medical imaging for position verification \cite{Moradi2020Robot-assistedSurgery,Song2022Real-timeDemonstration} in which the conventional US transducer can be used for simultaneous PA and US imaging. Robot-guided radical prostatectomy has been recognized as one of the best treatment options in patients with localized disease, while showing minimal long-term complications \cite{Roehl2004, Badani2007}.

To improve the SNR in PAI, temporal averaging is usually applied. This, however, lowers the imaging frame rate and increases the laser dose. Recently, a deep learning (DL) approach was taken by our group to mitigate this problem for PAI with the high-power Nd:YAG laser system \cite{Refaee2021DenoisingNetworks}, by training a Pix2Pix conditional generative adversarial network (cGAN) \cite{pix2pix2017}. This method aimed to denoise the pre-beamformed radio-frequency (RF) data, as well as removing sensor-specific artifacts, by creating a model that will translate a noisy image to a corresponding temporal-averaged image. A GAN consists of two main structures: the generator and the discriminator. The generator is trained to generate a denoised output based on a noisy input, whereas the discriminator is trained to distinguish that generated output from the reference dataset. However, low-energy LD-based PAI and image reconstruction from noisy data were not addressed in this paper.

DL has also been applied in PAI reconstruction algorithms \cite{Kim2020Deep-LearningSystem,Hauptmann2020DeepDirections,Lan2019Ki-GAN:Vivo,Lan2020Y-Net:Vivo,Guo2022AS-Net:Data,Waibel2018ReconstructionLearning,Reiter2017AData, Johnstonbaugh2020AMedium,Johnstonbaugh2019NovelLocalization, Allman2018PhotoacousticLearning}. The Delay-and-Sum (DAS) algorithm is widely acknowledged as an easy and simple reconstruction algorithm, but suffers from low image contrast and reconstruction artifacts \cite{Jeon2019}. In \cite{Lan2019Ki-GAN:Vivo,Lan2020Y-Net:Vivo,Guo2022AS-Net:Data}, different deep neural networks (DNNs) have been employed for PA tomography reconstruction. These models were trained on simulated PA signals and validated on both simulated PA signals and \textit{in vivo} experimental data. In \cite{Waibel2018ReconstructionLearning, Lan2019Ki-GAN:Vivo, Lan2020Y-Net:Vivo, Guo2022AS-Net:Data}, both the raw RF data and the DAS-reconstructed image are used as the model inputs. The paper of Waibel \textit{et al.} \cite{Waibel2018ReconstructionLearning} showed that using only raw RF data as the input was insufficient for good image reconstruction. These models, however, have not been applied in point source reconstruction, where different DL-based approaches are discussed. In \cite{Reiter2017AData, Johnstonbaugh2020AMedium}, point sources are recovered by acquiring numerical coordinates using convolutional neural networks (CNNs). In \cite{Waibel2018ReconstructionLearning, Lan2019Ki-GAN:Vivo, Lan2020Y-Net:Vivo, Guo2022AS-Net:Data, Reiter2017AData, Johnstonbaugh2020AMedium, Johnstonbaugh2019NovelLocalization, Allman2018PhotoacousticLearning}, all models are trained on simulated data, rather than experimental data, which can be attributed to the lack of availability of PAI data due to the limited clinical application.

In this study, we aimed to denoise pre-beamformed PAI data from a LD-based PA system that can mimic the behaviour of multi-frame temporal averaging by using a single frame. The option of employing a DNN for combined denoising and image reconstruction in LD-based PAI was also explored. Furthermore, as combining DL-based models trained for specific tasks has proven useful \cite{vanBoxtel2021HybridImages,Xu2017Dual-stageVasculopathy}, we will combine the denoising model and the reconstruction model. We will show that denoising and reconstruction are possible using a single frame of RF data. Our study brings novelty in several different ways:
\begin{itemize}\setlength\itemsep{0em}
    \item We use experimental data for training of our DL-based model rather than a large simulated dataset. This ensures that the model is trained and evaluated with accurate, real data with realistic SNR and noise distribution, which includes transducer artifacts and/or interference of signals.
    \item We employ our DL models to LD-based PAI, rather than PAI with high-power lasers. With the previously described benefits of LD over high-power lasers, LD may come close to matching the performance of high-power lasers through DL-based methods.
    \item We show that both denoising of pre-beamformed RF images and direct reconstruction are possible within real-time processing times.
    \item We employ a combined denoising-reconstruction strategy, which can improve on the reconstruction algorithm for very low-SNR inputs.
\end{itemize}

The remainder of this paper is structured as follows: in Section \ref{subsec:trainingdata}, the process of data acquisition is described, along with a description of the different datasets that were constructed from the acquired data. Section \ref{subsec:methods-algorithm} describes the denoising algorithms that were evaluated on the data and using the metrics outlined in Section \ref{subsec:methods-evaluation}.
Section \ref{subsec:methods-reconstruction} describes the algorithm that was used for automated reconstruction.
In Section \ref{sec:results}, our findings are described and these findings are discussed in Section \ref{sec:discussion}.

\section{Methods}\label{sec:method}
In the following sections, scalars are denoted by lowercase symbols and data matrices are denoted by bold uppercase symbols. A single sample from a data matrix is denoted by a regular uppercase with a superscript. We begin with describing the data acquisition process and the steps that were taken to construct the datasets for training and evaluation. Next, the denoising algorithms that were employed are described, as well as the training procedure. Methods for evaluation of those algorithms will be discussed after that. Finally, we discuss the methods used for automatic reconstruction using DL.

\subsection{RF dataset for training}\label{subsec:trainingdata}

\subsubsection{Data acquisition}
\begin{figure}[tb]
    \centering
    \includegraphics[width=0.85\linewidth]{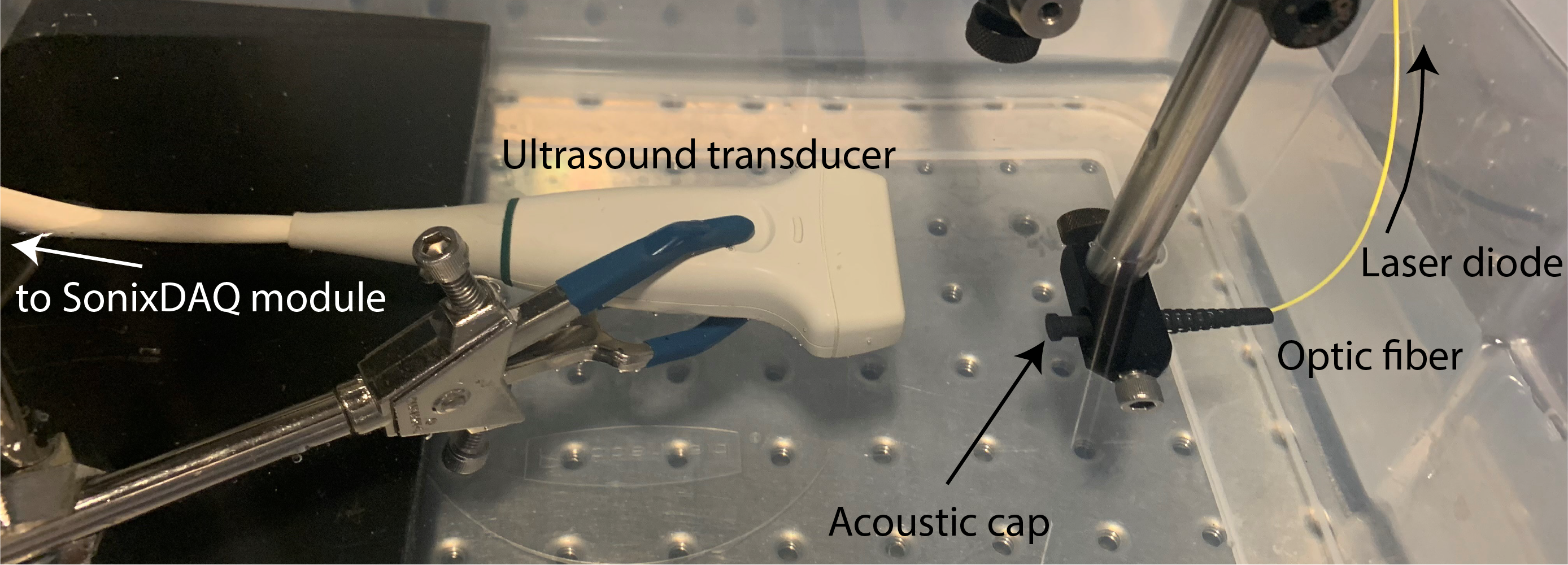}
    \caption{Experimental setup for acquisition of training and validation data with the LD PA source. The whole setup is immersed in water to reduce impedance effects.}
    \label{fig:exp-setup}
\end{figure}

In our study, the PA signal was induced by a pulsed laser diode (QPhotonics QSP-915-20, MI, USA), operating at a wavelength of 915 nm with a power of 20W driven by a pulsed laser driver (PicoLAS LDP-V 50-100 V3.3, Germany) to generate 100 ns laser pulses. Acoustic waves were generated using a black plastic cap around the laser diode output.
These acoustic waves were measured using a 128-element linear transducer (Ultrasonix L14-5), and an Ultrasonix SonixDAQ module was used to record the RF data at a frequency of 40 MHz. The experimental setup for these acquisitions is visualized in Figure \ref{fig:exp-setup}. The entire setup was immersed in water, to mitigate impedance effects. Acquisitions were made at different axial distances between the LD output and transducer. With the LD in-plane relative to the transducer, axial distances of 3, 4, 6 and 7 cm were used. With the LD out-of-plane of the transducer, the distance was set to 3 and 4 cm. For each distance, the LD was moved in the lateral direction to capture 20 random lateral point sources. 16 of these 20 acquisitions were assigned for training, with the remaining 4 being assigned for intermediate model validation and for optimization of the model hyperparameters on unseen data. This should have resulted in 96 and 24 datasets in training and validation sets respectively, however, after removing some duplicated training datasets, 84 unique datasets remained in the training set.

\subsubsection{Reference dataset}
A single frame of RF data is described by a $(n_t,n_e)$ matrix $\boldsymbol{X_\text{1f}}$, where $n_e$ is the number of elements and $n_t$ is the number of recorded time points. For every dataset, $n_f$ frames are recorded. To obtain a larger variety of different strengths and noise patterns, the 3- and 5-frame averages were also calculated, denoted by $\boldsymbol{X_\text{3f}}$ and $\boldsymbol{X_\text{5f}}$, respectively. Moreover, temporal averaging was applied over all $n_f$ frames to create a reference dataset $\boldsymbol{X_\text{r}}$, with the same shape $(n_t,n_e)$. A 1-D median filter of size 5 was applied in the temporal direction of the reference RF data, to resolve some artifacts incurred by the SonixDAQ machine. In this study, $n_e = 128$ elements, $n_t = 2000$ time points, and $n_f = 260$ frames.

Despite the temporal averaging and the spatial median filtering, image artifacts remained. In the training dataset of 84 images, we segmented the background manually. For every pixel segmented as `background', its pixel value was replaced by the median value of all background pixels in the image. This segmented reference image is denoted by $\boldsymbol{X_\text{s}}$ and has the same size of $(n_t,n_e)$.

\begin{figure}[bt]
    \centering
    \begin{subfigure}{0.14\linewidth}
    \includegraphics[width=\linewidth]{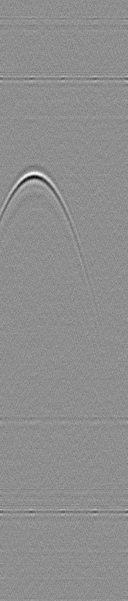}
    \caption*{Single frame ($\boldsymbol{X_{\text{1f}}}$)}
    \end{subfigure} \hspace{2mm}
    \begin{subfigure}{0.14\linewidth}
    \includegraphics[width=\linewidth]{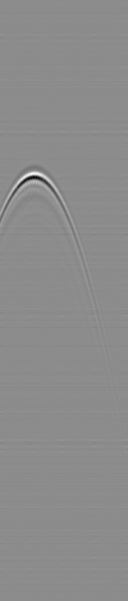}
    \caption*{Reference image ($\boldsymbol{X_{\text{r}}}$)}
    \end{subfigure} \hspace{2mm}
    \begin{subfigure}{0.14\linewidth}
    \includegraphics[width=\linewidth]{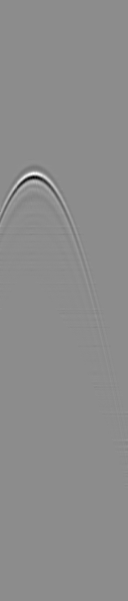}
    \caption*{Segmented reference image ($\boldsymbol{X_{\text{s}}}$)}
    \end{subfigure} \hspace{2mm}
    \begin{subfigure}{0.14\linewidth}
    \includegraphics[width=\linewidth]{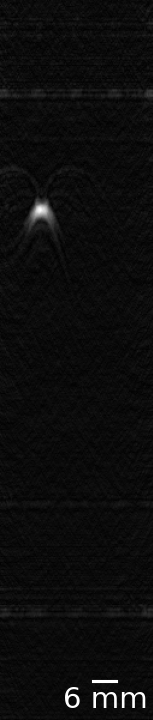}
    \caption*{DAS from single frame}
    \end{subfigure} \hspace{2mm}
    \begin{subfigure}{0.14\linewidth}
    \includegraphics[width=\linewidth]{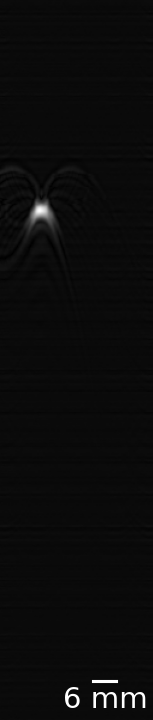}
    \caption*{DAS from $n_f$ frames ($\boldsymbol{X_{\text{Dr}}}$)}
    \end{subfigure} \hspace{2mm}
    \begin{subfigure}{0.14\linewidth}
    \includegraphics[width=\linewidth]{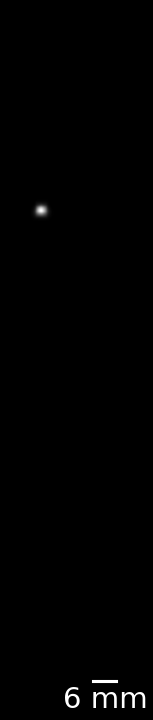}
    \caption*{Exact reconstruction ($\boldsymbol{X_{\text{De}}}$)}
    \end{subfigure} \hspace{2mm}
    
    \caption{Samples from different data types. Single-frame DAS was calculated from $\boldsymbol{X_{\text{1f}}}$, $\boldsymbol{X_{\text{Dr}}}$ was calculated from $\boldsymbol{X_{\text{r}}}$, and then, $\boldsymbol{X_{\text{De}}}$ was obtained from $\boldsymbol{X_{\text{Dr}}}$. Note that $\boldsymbol{X_{\text{s}}}$ and $\boldsymbol{X_{\text{De}}}$ are only obtained for samples in the test set.}
    \label{fig:data_subsets}
\end{figure}

To generate reference images to train the reconstruction model, the DAS reconstruction was calculated from $\boldsymbol{X_{\text{r}}}$, yielding reconstructed image datasets $\boldsymbol{X_{\text{Dr}}}$. The DAS algorithm was set up so that these reconstructions had the same size as the RF data, i.e., $(n_t,n_e)$. However, this DAS reconstruction is not very specific and contains some artifacts. To get as close to a `true reconstruction' as possible, and given that the training dataset size was small enough, the location of the point source was selected manually in $\boldsymbol{X_{\text{Dr}}}$. Then, this selected point was isolated by applying a 2-D Gaussian distribution mask around the selected point on the reference DAS reconstruction, where $\sigma_{\text{Gauss}} = 0.3\ \text{[mm]}$, equivalent to the size of the point source that was measured. This results in a single point source on an empty background, and is here denoted by $\boldsymbol{X_{\text{De}}}$ of the same size. In Figure \ref{fig:data_subsets}, the different data subsets are visualized.

\subsubsection{Dataset for model training}
For training a DL-based model, having limited data, 84 images in our example, often causes the model to not generalize well to unseen data. Especially on larger models and GANs, overfitting is a big problem \cite{Ying2019AnSolutions}. Therefore, we present different methods to increase the number of training samples.

First, $n_f$ frames per spot were collected. With $n_f = 260$ frames acquired per dataset in our example, it is theoretically possible to use all those frames. Due to the similarity of the systemic noise patterns of single frames of the same experiment, to include the random noise, and to optimize the memory usage, here, 5\% of all available frames were used. The same method was applied for $\boldsymbol{X_{\text{3f}}}$ and $\boldsymbol{X_{\text{5f}}}$, which also both used 5\% of the available frames. These three extended datasets were combined into dataset $\boldsymbol{X_{\text{F}}}$, yielding a total number of 1596 images in our example:

\begin{equation}
    \left[\boldsymbol{X_\text{F}}\right]_{1596 \times 2000 \times 128} = \left[ \left[\boldsymbol{X}_\text{1f}\right]_{1092 \times 2000 \times 128} \left[\boldsymbol{X}_\text{3f}\right]_{336 \times 2000 \times 128} \left[\boldsymbol{X}_\text{5f}\right]_{168 \times 2000 \times 128} \right]
\end{equation}

\begin{figure}[bt]
    \centering
    \includegraphics[width=0.65\linewidth]{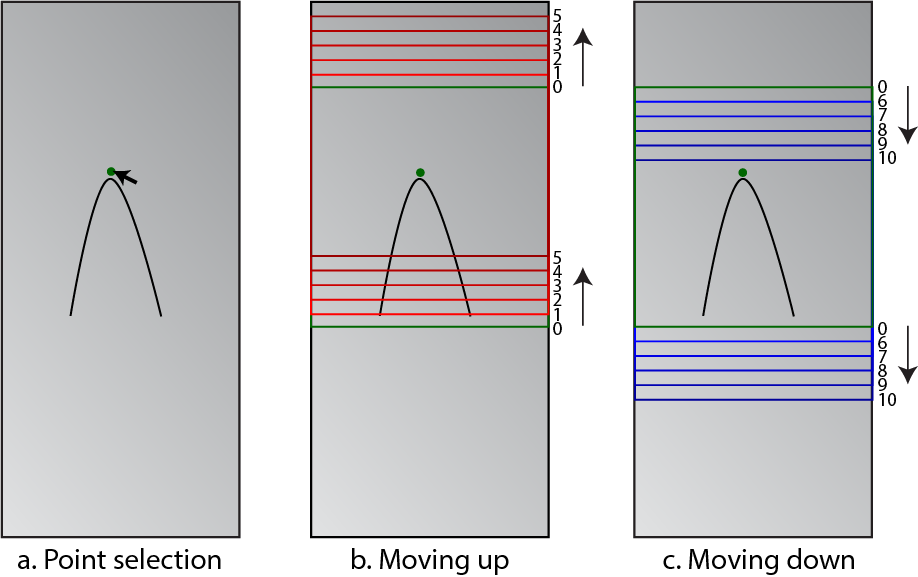}
    \caption{Illustration on the moving-window patch selection method. (a) Manually select the top of the typical signal-parabola; (b) A patch will be defined around the selected point, and next, the patch location will move up by a predefined step size for a predefined number of steps; (c) From the selected point, this process will be repeated, but now, moving down from the selected point.}
    \label{fig:moving_window}
\end{figure}

Next, instead of using the RF matrices shaped $(n_t,n_e)$, patches of $(n_e, n_e)$ pixels were extracted by selecting patches containing signal, and separately, patches containing noise. This was done to acquire a balanced dataset, resulting in an approximately equal number of noisy and signal patches. Given the limited number of experimental data, to get the signal at different positions and depths within the patches, a `moving window' method was used, which would use the manually selected patch location and move the patch up and down along the time-axis, as illustrated by Figure \ref{fig:moving_window}. For each training dataset, 21 signal patches and 21 noise patches were selected using the moving window method by going 10 steps up and down from the selected patch, further increasing the training dataset size by 42 times.
From here on, datasets containing patches with shape $(n_e,n_e)$ are denoted by the symbol $\boldsymbol{Z}$. A subscription will indicate whether the patches originate from a single frame (F), the reference image (r), segmented image (s) or exact reconstruction (De).

\subsection{Denoising algorithms}\label{subsec:methods-algorithm}

\subsubsection{Pix2Pix GAN}\label{subsubsec:methods-pix2pix}
The Pix2Pix GAN model was employed, as it was previously shown that this model can perform well for denoising PA data acquired with high-power lasers \cite{Refaee2021DenoisingNetworks}. The Pix2Pix model is trained to learn a mapping, to convert an input image to an output image. The training objective of the Pix2Pix model can be described as:

\begin{align}\label{eq:losses1}
    L_\text{GAN} &= \mathbb{E}_{\boldsymbol{Z_\text{F}},\boldsymbol{Z_\text{r}}} \left[\log D\left(\boldsymbol{Z_\text{F}},\boldsymbol{Z_\text{r}} \right)   \right] + \mathbb{E}_{\boldsymbol{Z_\text{F}},\boldsymbol{W}} \left[\log\left(1-D\left(\boldsymbol{Z_\text{F}},G\left( \boldsymbol{Z_\text{F}},\boldsymbol{W} \right) \right) \right) \right] \\[2mm]\label{eq:losses2}
    L_\text{L1} &= \mathbb{E}_{\boldsymbol{Z_\text{F}},\boldsymbol{Z_\text{r}},\boldsymbol{W}} \left[ \lVert{\boldsymbol{Z_\text{r}} - G\left(\boldsymbol{Z_\text{F}},\boldsymbol{W}  \right)}\rVert_1 \right] \\[2mm]\label{eq:losses3}
    G^* &= \arg \min_G \max_D \left\{ L_\text{GAN} + \lambda_1 L_\text{L1}  \right\}
\end{align}

Here, $L_\text{GAN}$ is the GAN loss, $L_\text{L1}$ is the L1 distance norm, $\mathbb{E}$ is the expectation value, and $\boldsymbol{W}$ is a random noise vector with the same shape as $\boldsymbol{Z}$, i.e., $(n_e, n_e)$. By controlling the parameter $\lambda_1$, the regularization using the L1 norm can be changed \cite{pix2pix2017}. In case that the model was trained on segmented data as output, $\boldsymbol{Z_\text{r}}$ will change to $\boldsymbol{Z_\text{s}}$ in the Equations \ref{eq:losses1} and \ref{eq:losses2}.

\begin{figure}
    \centering
    \includegraphics[width=0.8\linewidth]{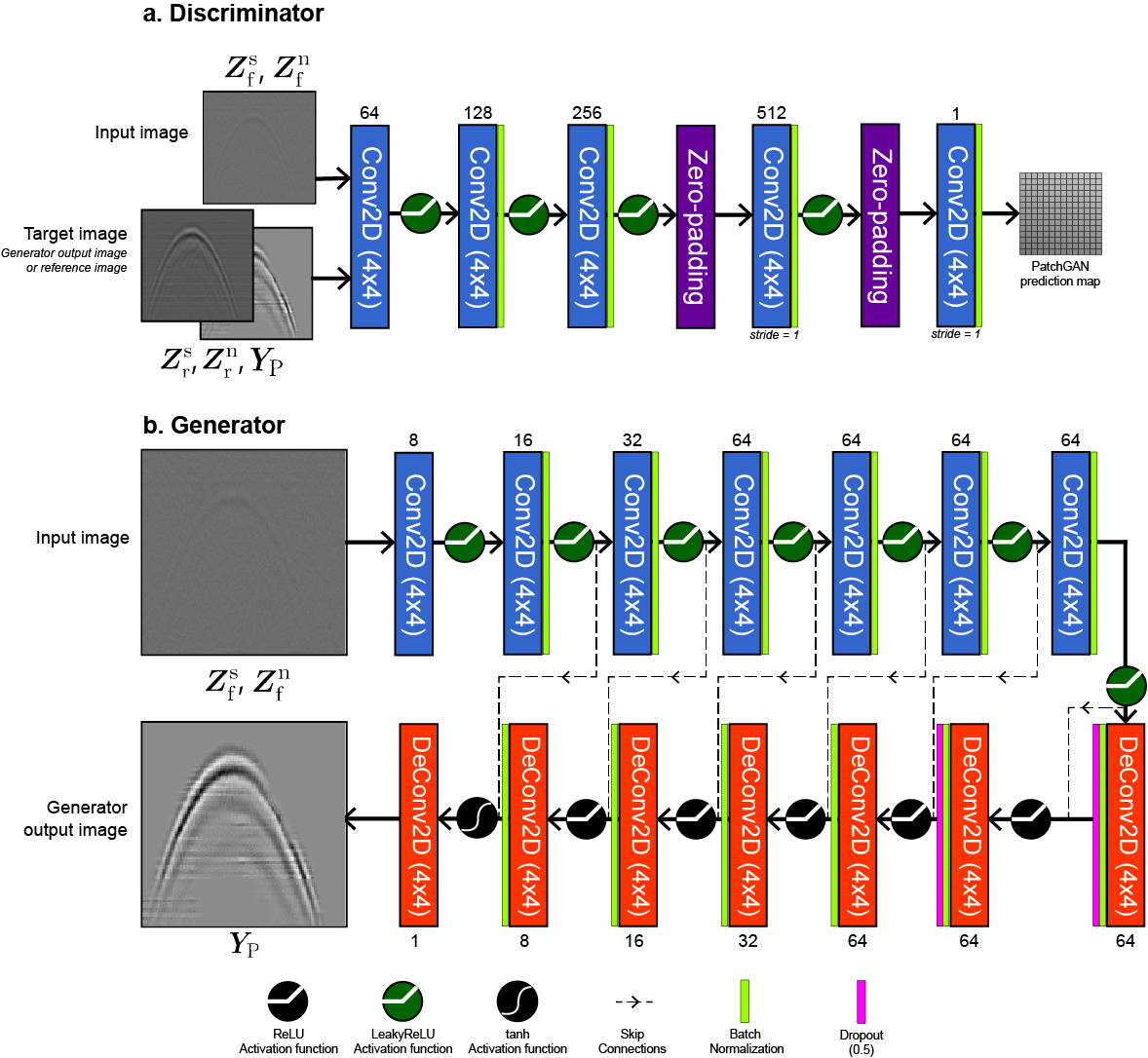}
    \caption{(a) Discriminator of the Pix2Pix cGAN model. (b) Generator of the Pix2Pix cGAN model.}
    \label{fig:modelarch}
\end{figure}

The model architectures for both generator and discriminator are shown in Figure \ref{fig:modelarch}. In the Pix2Pix model, the generator used was an adaption to the U-Net architecture \cite{Ronneberger2015U-NetSegmentation}. Compared to many other previous studies \cite{Refaee2021DenoisingNetworks,Chi2020denoising,Han2018FramingCT}, the U-Net is much smaller in terms of the number of model parameters, using only 458,785 parameters compared to 10,471,425 parameters in the U-Net model as used in \cite{Refaee2021DenoisingNetworks}. A smaller model has an advantage in computation time and reduces the risk for overfitting.

The discriminator uses a simple CNN layout named PatchGAN \cite{Isola2017Image-to-imageNetworks}, which consists of five hidden layers. The generator is given noisy inputs, which generates a prediction for a denoised image. This prediction, along with the noisy and the reference frames, would then be used as input for the discriminator, which would give a binary prediction as output: a `0' as prediction would imply a `fake' image (created by the generator), whereas a `1' as prediction would imply that the discriminator thinks that this is a `real' reference image. 

The generator and discriminator were optimized during 5000 epochs using the Adam optimizer, with an initial learning rate of $\alpha = 2\cdot 10^{-4}$, L1 regularization parameter $\lambda_\text{1}=10^3$, a batch size of $n_\text{batch} = 32$, and minibatches of 1024 images per epoch. These values were obtained by performing a small hyperparameter search. The datasets containing signal and noise patches from single frames, 3- and 5-frame averages were used as input to the model. Corresponding patches, generated from $\boldsymbol{X_\text{r}}$ and $\boldsymbol{X_\text{s}}$ were used as target output of the model in two separate experiments. Training was done on a NVIDIA Tesla V100 GPU. 

\subsubsection{Pix2Pix-Residual GAN}
The Pix2Pix-Residual GAN shows a lot of similarities to the Pix2Pix GAN, however, the `default' U-Net generator was here replaced with a more advanced U-Net-like architecture, which contains more residual connections between hidden layers. This model was introduced for image denoising by Gurolla-Ramos \textit{et al.} \cite{Gurrola-Ramos2021ADenoising}, but has, to our knowledge, not yet been used in a GAN approach. For the exact architecture, we refer to \cite{Gurrola-Ramos2021ADenoising}. The residual U-Net is a much larger model in terms of model parameters due to the residual connections. Here, the residual U-Net generator had a total of 3,938,553 model parameters, as the number of filters in the first layer was changed from $f=128$ to $f=8$, and still increasing with a factor 2 for the following layers in the encoder. The PatchGAN discriminator from the original Pix2Pix model was retained in this GAN.

The generator and discriminator were optimized during 5000 epochs using the Adam optimizer, with an initial learning rate of $\alpha = 2\cdot 10^{-4}$, L1 regularization parameter $\lambda_\text{1}=10^3$, a batch size of $n_\text{batch} = 32$, and minibatches of 1024 images per epoch. Datasets containing patches extracted from $\boldsymbol{X_\text{F}}$ were used as input to the model, and again, corresponding patches originating from $\boldsymbol{X_\text{r}}$ and $\boldsymbol{X_\text{s}}$ were used as target output of the model in two separately trained models.

\subsection{Evaluation of denoising algorithm}\label{subsec:methods-evaluation}
\begin{figure}[tb]
    \centering
    \includegraphics[width=0.9\linewidth]{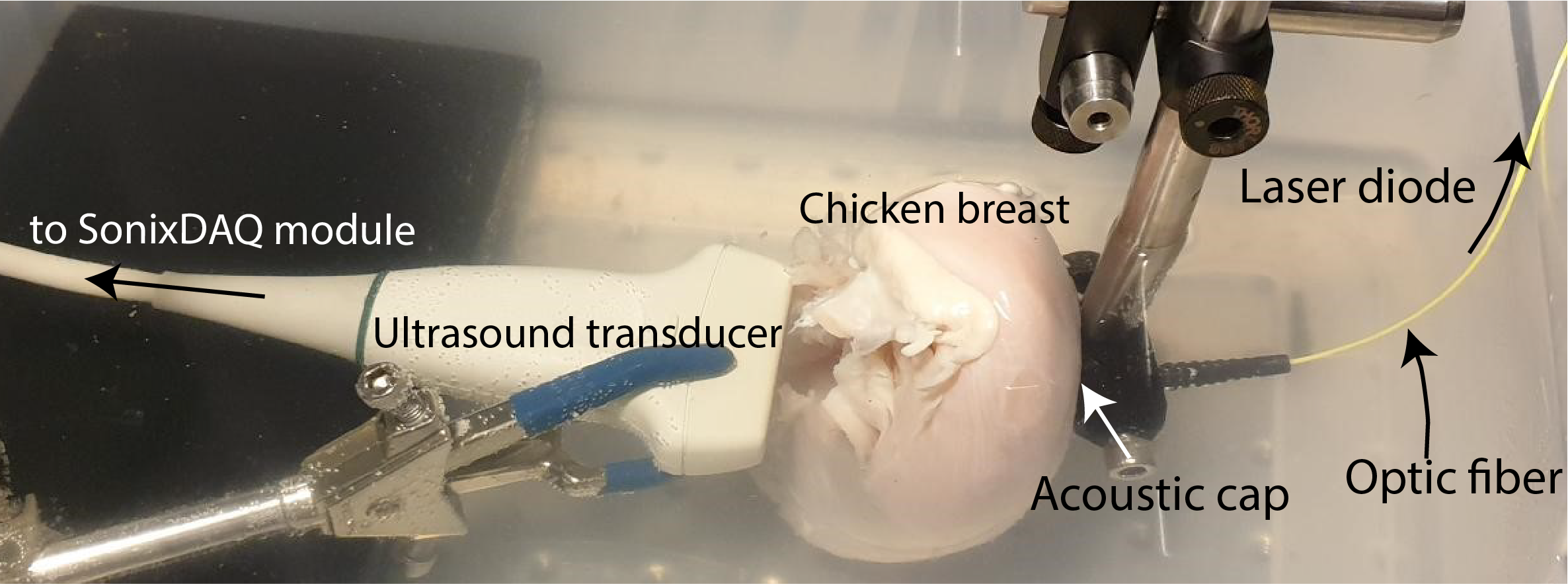}
    \caption{Experimental setup for acquisition of test data with the LD PA source. The whole setup is immersed in water to reduce impedance effects.}
    \label{fig:exp-setup-testdata}
\end{figure}

For evaluation of our methods, new RF data was collected. The experimental setup is shown in Figure \ref{fig:exp-setup-testdata}. The entire setup was immersed in water to reduce impedance effects. Measurements were done with the LD output at two random axial distances with \textit{ex vivo} tissue, a chicken breast sample, between the transducer and LD output. The LD was moved in the lateral direction, collecting 10 random lateral point sources yielding a total of 20 datasets.

Predictions from the denoising algorithm were evaluated based on visual evaluation and two metrics. Visual evaluation can prove useful when specific patterns or artifacts occur in the model predictions. As all models were trained on square patches, the evaluation data will be split into patches as well, and reconstructed back to a single image after model prediction. Patches were extracted overlapping with the previous and next patch, covering the whole image. In other words, after extracting a patch, the next patch is located 64 pixels downwards in our case. 

Here, model predictions will be denoted by $\boldsymbol{\hat{X}}$. The metrics selected for evaluation of the model predictions are Mean Squared Error (MSE; Eq. \ref{eq:MSE}) and Structural Similarity Index Measure (SSIM; Eq. \ref{eq:SSIM}) \cite{Wang2004ImageSimilarity}:

\begin{equation} \label{eq:MSE}
    \mathrm{MSE} = \frac{1}{n_i} \sum^{n_i}_{i=1} \left({x^i_{\text{r}}} - \hat{x}^i \right)^2
\end{equation}

\begin{equation}\label{eq:SSIM}
    \mathrm{SSIM} = \frac{(2 \mu_{\hat{x}} \mu_{\text{r}} + c_1) (2 \sigma_{{x}_\text{r} {\hat{x}}} + c_2)}{ (\mu_{\hat{x}}^2 + \mu_{\text{r}}^2 + c_1) (\sigma_{\hat{x}}^2 + \sigma_{\text{r}}^2 + c_2) }
\end{equation}

\begin{equation}
    \sigma_{{x}_\text{r},{\hat{x}}} = \frac{1}{n_j-1} \sum^{n_j}_{j=1} ({x^i_{\text{r}}}-\mu_{\text{r}}) ({\hat{x}^i}-\mu_{\hat{x}}) 
\end{equation}

with $x_\text{r}^i$ and $\hat{x}^i$ being single pixels in $\boldsymbol{X_\text{r}}$ and $\hat{\boldsymbol{X}}$, respectively, $n_i$ being the number of pixels in an image, $\mu$ the mean signal intensity, $\sigma$ the standard deviation of the signal, and $\sigma_{{x}_\text{r},\hat{x}}$ the correlation coefficient of $X_\text{r}$ and $\hat{\boldsymbol{X}}$. $c_1$ and $c_2$ are small constants to avoid ill-defined values, and are set to $c_1 = 0.01$ and $c_2 = 0.03$, matching the values defined in \cite{Wang2004ImageSimilarity}.

Next to these metrics, the evaluation time is measured when predicting on a modern laptop, using the CPU (Intel i7-11800H) or GPU (NVIDIA RTX 3060 Mobile), to assess whether a real-time application of these models would be viable.

\subsection{Reconstruction algorithm} \label{subsec:methods-reconstruction}
Different DL methods have been applied for PA reconstruction \cite{Kim2020Deep-LearningSystem,Hauptmann2020DeepDirections,Lan2019Ki-GAN:Vivo,Lan2020Y-Net:Vivo,Guo2022AS-Net:Data,Waibel2018ReconstructionLearning,Reiter2017AData, Johnstonbaugh2020AMedium,Johnstonbaugh2019NovelLocalization, Allman2018PhotoacousticLearning}. Here, we propose a direct method for reconstruction from noisy RF data when measuring point sources. For this, we used the Pix2Pix model with the same hyperparameters as described in Section \ref{subsubsec:methods-pix2pix}. However, instead of using reference or segmented images, the target output was changed to the exact reconstruction. This adapts the loss functions to:
\begin{align}
    L_\text{GAN} &= \mathbb{E}_{\boldsymbol{Z_\text{F}},\boldsymbol{Z_\text{De}}} \left[\log D\left(\boldsymbol{Z_\text{F}},\boldsymbol{Z_\text{De}} \right)   \right] + \mathbb{E}_{\boldsymbol{Z_\text{F}},\boldsymbol{W}} \left[\log\left(1-D\left(\boldsymbol{Z_\text{F}},G\left( \boldsymbol{Z_\text{F}},\boldsymbol{W} \right) \right) \right) \right] \\[2mm]
    L_\text{L1} &= \mathbb{E}_{\boldsymbol{Z_\text{F}},\boldsymbol{Z_\text{De}},\boldsymbol{W}} \left[ \lVert{\boldsymbol{Z_\text{De}} - G\left(\boldsymbol{Z_\text{F}},\boldsymbol{W}  \right)}\rVert_1 \right] \\[2mm]
    G^* &= \arg \min_G \max_D \left\{ L_\text{GAN} + \lambda_1 L_\text{L1}  \right\}
\end{align}

Here, the generator and discriminator were optimized during 5000 epochs using the Adam optimizer, using an initial learning rate of $\alpha = 2\cdot 10^{-4}$, L1 regularization parameter $\lambda_\text{1}=10^3$, a batch size of $n_\text{batch} = 32$, and minibatches of 1024 images per epoch.

We also explored the possibility of combined denoising and reconstruction, which we will here call Dual-GAN. A single denoising GAN was selected based on its performance, and was given single frames as the input data. The output from this denoising GAN was then given as input data to the reconstruction GAN, yielding the Dual-GAN output.

\section{Results}\label{sec:results}
In this section, we show the results of our findings. First, we describe the results of the denoising algorithm, and next, we show the results of the reconstruction algorithm. All results that are shown here are based on samples from the test set, i.e., data that was previously unseen by any of the models.

\subsection{Denoising algorithm}
In Table \ref{tab:metrics-denoise}, two different metrics are shown for different denoising methods. In all metric calculations, samples from the test set are used, as they are previously unseen by the models, and reference images were used as the ground truth. For the cases where the models were trained on segmented images as the target output, the metric values may not be entirely accurate, since the `true reference' for this model should have an empty background. However, since these segmented images were only available for the training data, the reference images ($\boldsymbol{X_\text{r}}$) are used for these calculations.

\begin{figure}[!ht]
    \centering
    \begin{subfigure}{0.275\linewidth}
    \includegraphics[width=\linewidth]{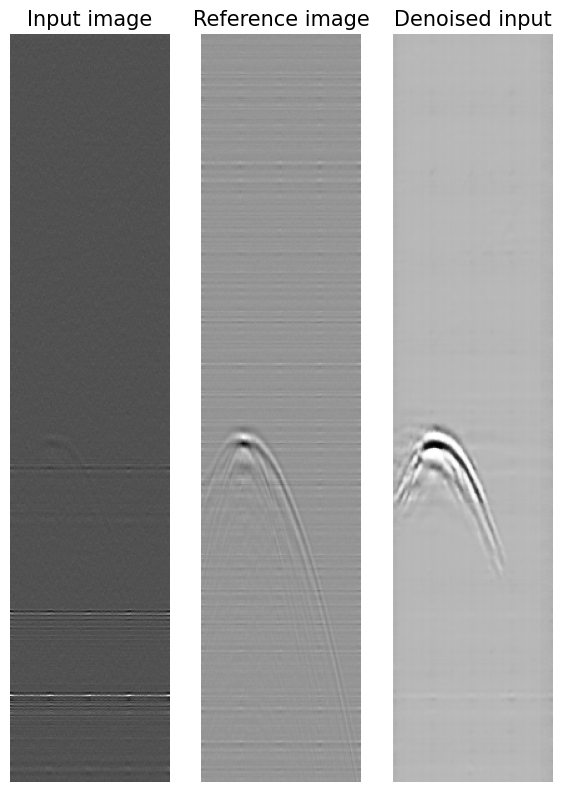}\hfill
    \caption{}
    \end{subfigure}\hspace{0.5cm}
    \begin{subfigure}{0.275\linewidth}
    \includegraphics[width=\linewidth]{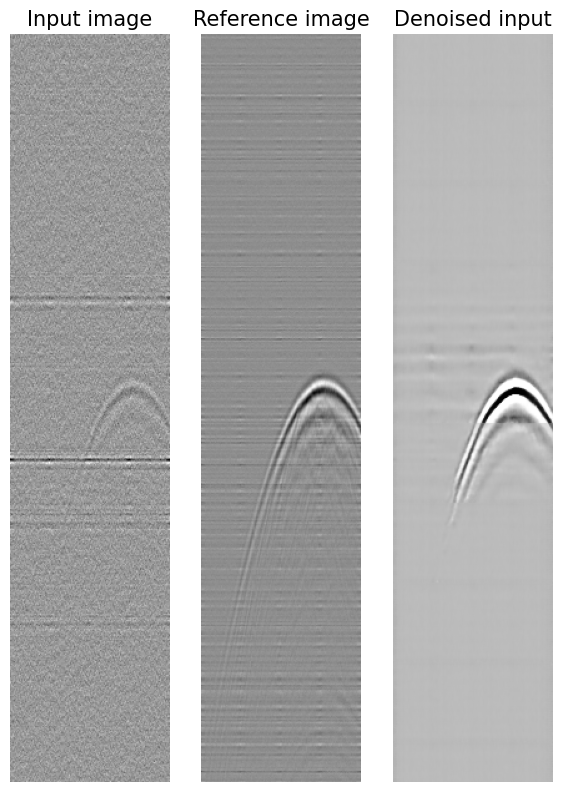}\hfill
    \caption{}
    \end{subfigure}\hspace{0.5cm}
    \begin{subfigure}{0.275\linewidth}
    \includegraphics[width=\linewidth]{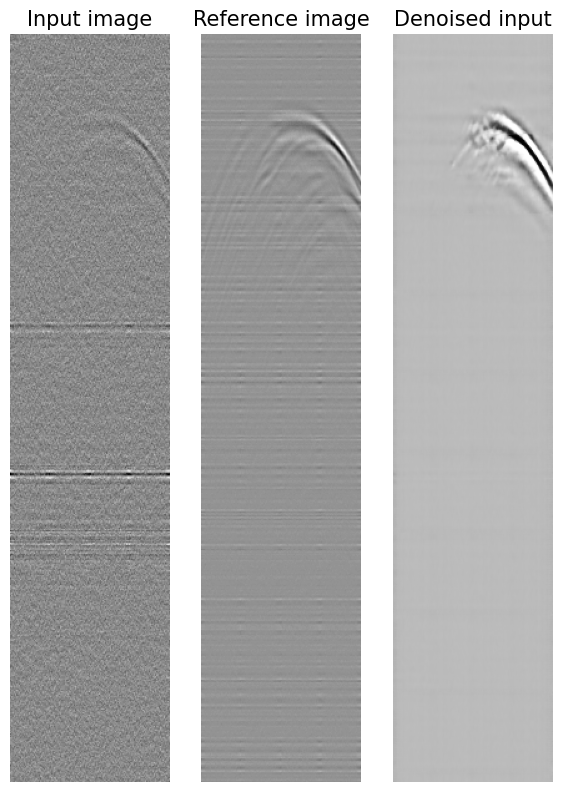}\hfill
    \caption{}
    \end{subfigure}
    \vspace{-0mm}
    \caption{Predictions of the Pix2Pix model, trained with reference targets.}
    \label{fig:denoise-p2p}
\end{figure}

\begin{figure}[!ht]
    \centering
    \begin{subfigure}{0.275\linewidth}
    \includegraphics[width=\linewidth]{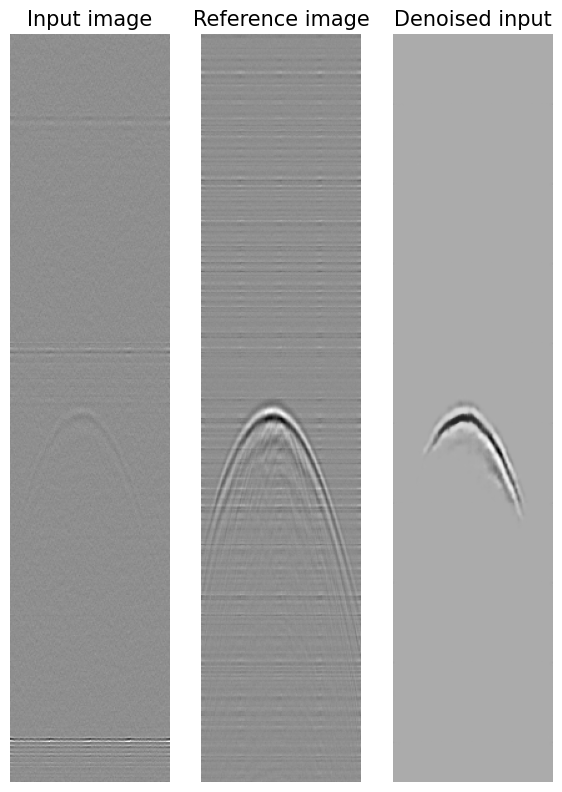}\hfill
    \caption{}
    \end{subfigure}\hspace{0.5cm}
    \begin{subfigure}{0.275\linewidth}
    \includegraphics[width=\linewidth]{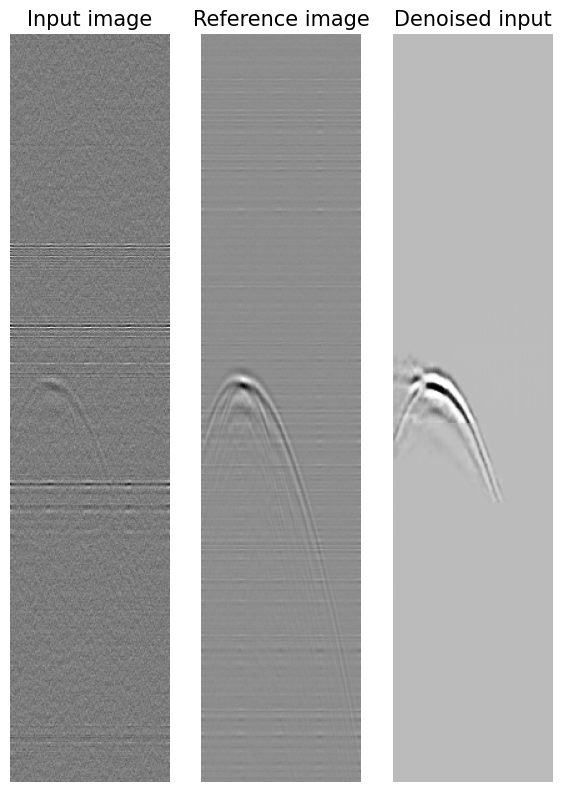}\hfill
    \caption{}
    \end{subfigure}\hspace{0.5cm}
    \begin{subfigure}{0.275\linewidth}
    \includegraphics[width=\linewidth]{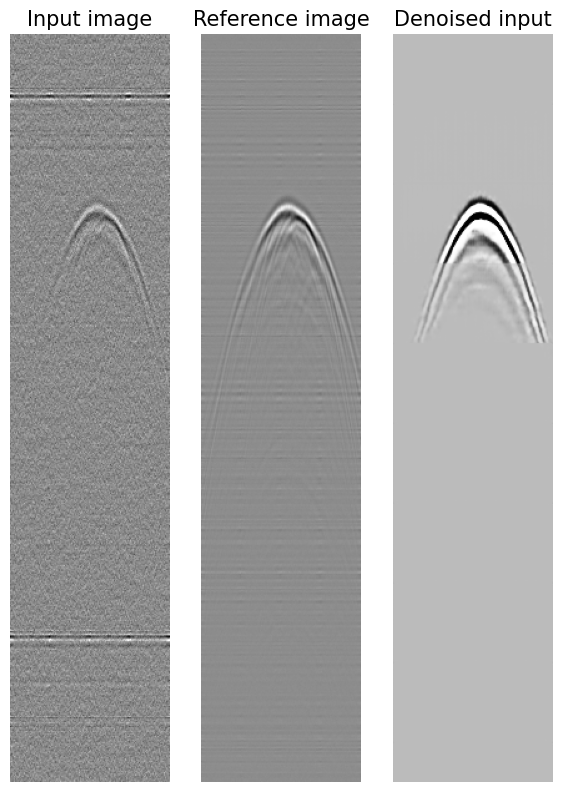}\hfill
    \caption{}
    \end{subfigure}
    \vspace{-0mm}
    \caption{Predictions of the Pix2Pix model, trained with segmented targets.}
    \label{fig:denoise-p2p-noBG}
\end{figure}

\begin{figure}[!ht]
    \centering
    \begin{subfigure}{0.275\linewidth}
    \includegraphics[width=\linewidth]{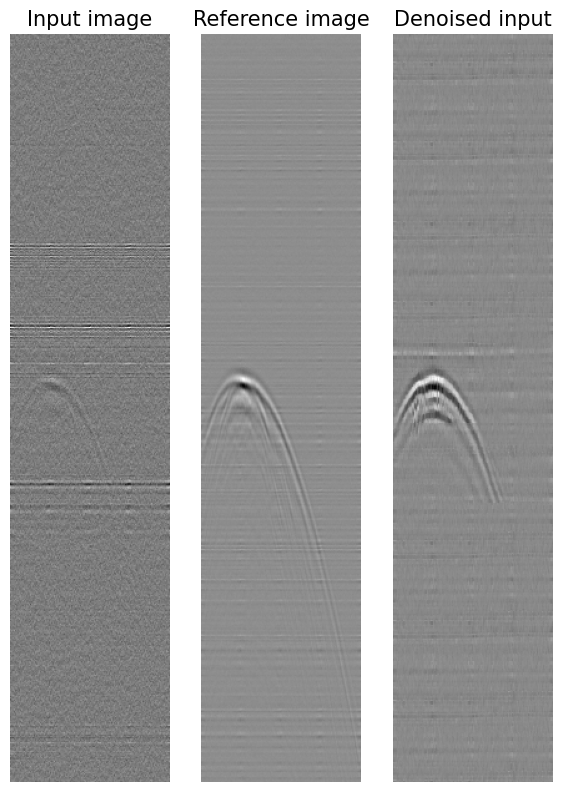}\hfill
    \caption{}
    \end{subfigure}\hspace{0.5cm}
    \begin{subfigure}{0.275\linewidth}
    \includegraphics[width=\linewidth]{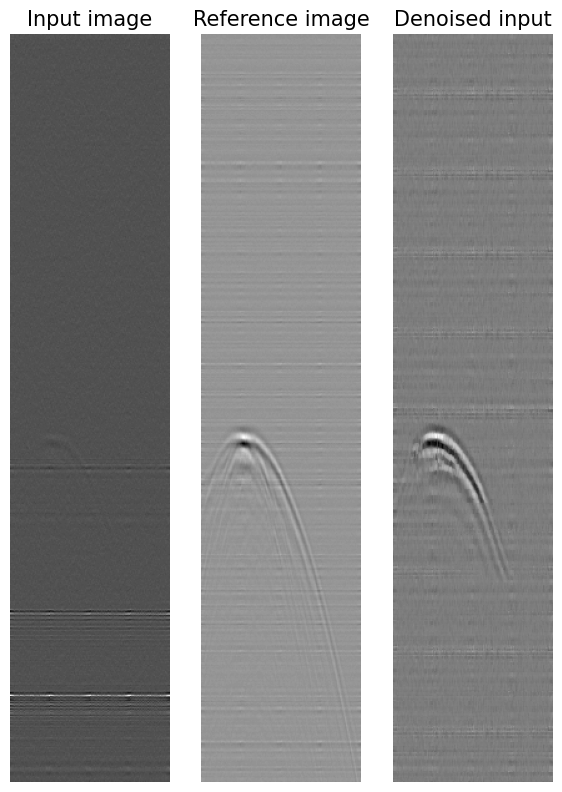}\hfill
    \caption{}
    \end{subfigure}\hspace{0.5cm}
    \begin{subfigure}{0.275\linewidth}
    \includegraphics[width=\linewidth]{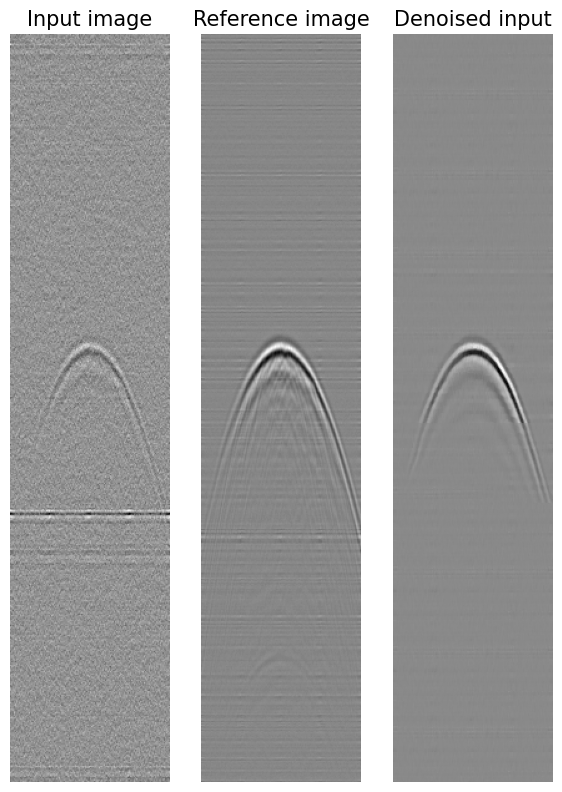}\hfill
    \caption{}
    \end{subfigure}
    \vspace{-0mm}
    \caption{Predictions of the Pix2Pix-Residual model, trained with reference targets.}
    \label{fig:denoise-p2pr}
\end{figure}

\begin{figure}[!ht]
    \centering
    \begin{subfigure}{0.275\linewidth}
    \includegraphics[width=\linewidth]{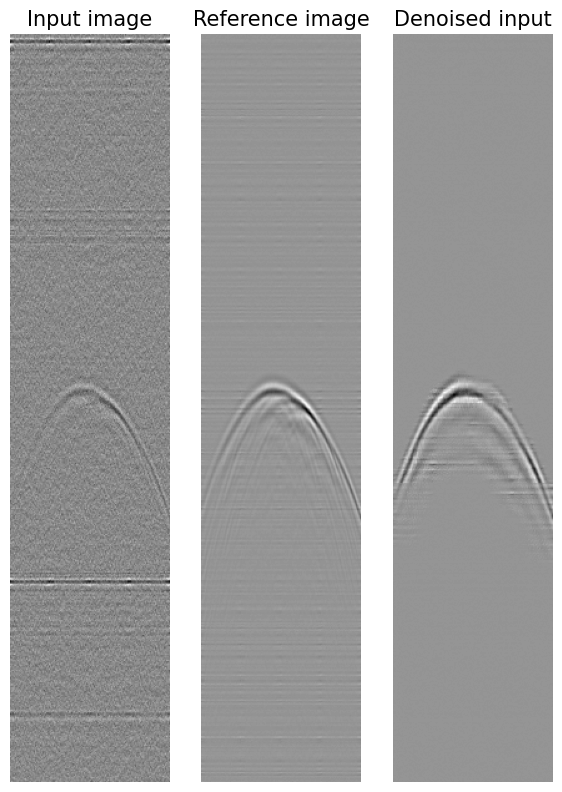}\hfill
    \caption{}
    \end{subfigure}\hspace{0.5cm}
    \begin{subfigure}{0.275\linewidth}
    \includegraphics[width=\linewidth]{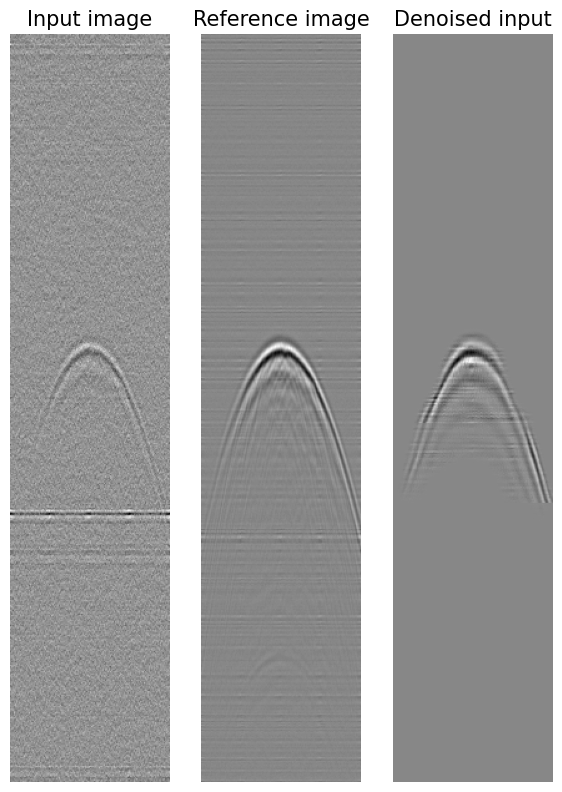}\hfill
    \caption{}
    \end{subfigure}\hspace{0.5cm}
    \begin{subfigure}{0.275\linewidth}
    \includegraphics[width=\linewidth]{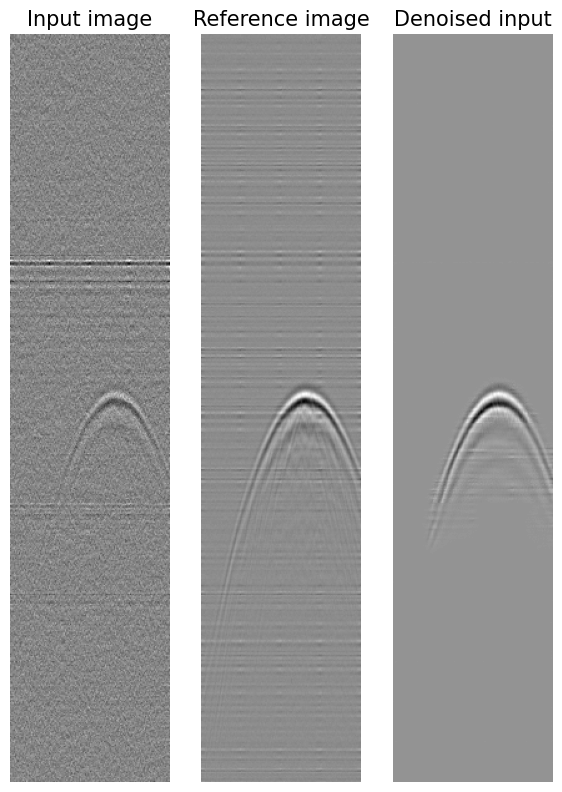}\hfill
    \caption{}
    \end{subfigure}
    \vspace{-0mm}
    \caption{Predictions of the Pix2Pix-Residual model, trained with segmented targets.}
    \label{fig:denoise-p2pr-noBG}
\end{figure}

In Figure \ref{fig:denoise-p2p}, some predicted samples from the Pix2Pix model trained on reference images are shown, along with the corresponding input frame and actual reference image. Figure \ref{fig:denoise-p2p-noBG} shows samples of the Pix2Pix model trained on segmented data. Similarly, in Figures \ref{fig:denoise-p2pr} and \ref{fig:denoise-p2pr-noBG}, results for the Pix2Pix-Residual model are shown when using reference data and segmented data as target outputs, respectively. Again, it should be noted that the ``Reference image'' in the previous figures is always the reference image ($\boldsymbol{X_\text{r}}$), since no segmented data is available for the test set.

\begin{table}[!t]
    \centering
    \caption{MSE and SSIM metrics for different denoising methods compared to the values for noisy data, as well as processing times on CPU and GPU.}
    \vspace{-3mm}
    \begin{tabular}{m{3.2cm} | m{1.9cm} m{1.9cm} | m{2.15cm} m{2.15cm}}
        \hline
        \textbf{Denoising method}   & \textbf{MSE} & \textbf{SSIM} & \textbf{Evaluation\newline time CPU [ms]} & \textbf{Evaluation\newline time GPU [ms]}  \\ 
        \hline 
        Noisy data       & 0.018 \textpm\ 0.013          & 0.744 \textpm\ 0.139          & -                    & -                \\ 
        10 frames averaged              & 0.021 \textpm\ 0.013          & 0.892 \textpm\ 0.028          & -                    & -                \\ 
        20 frames averaged              & 0.017 \textpm\ 0.010          & 0.883 \textpm\ 0.031          & -                    & -                \\ 
        Pix2Pix trained with reference targets       & \textbf{0.014 \textpm\ 0.004} & 0.914 \textpm\ 0.014          & 47.3 \textpm\ 1.8    & 34.9 \textpm\ 3.7 \\
        Pix2Pix trained with segmented targets      & 0.015 \textpm\ 0.005          & \textbf{0.926 \textpm\ 0.017} & 46.3 \textpm\ 1.5    & 35.4 \textpm\ 4.0 \\
        Pix2Pix-Residual trained with reference targets       & 0.047 \textpm\ 0.039          & 0.579 \textpm\ 0.020          & 893.0 \textpm\ 19.1  & 69.3 \textpm\ 4.9 \\
        Pix2Pix-Residual trained with segmented targets      & 0.021 \textpm\ 0.005          & 0.914 \textpm\ 0.017          & 874.4 \textpm\ 18.9  & 71.2 \textpm\ 5.1  \\ 
        \hline
    \end{tabular}
    \label{tab:metrics-denoise}
\end{table}

\subsection{Reconstruction algorithm}

\begin{figure}[!t]
    \centering
    \begin{subfigure}{\linewidth}
    \includegraphics[width=\linewidth]{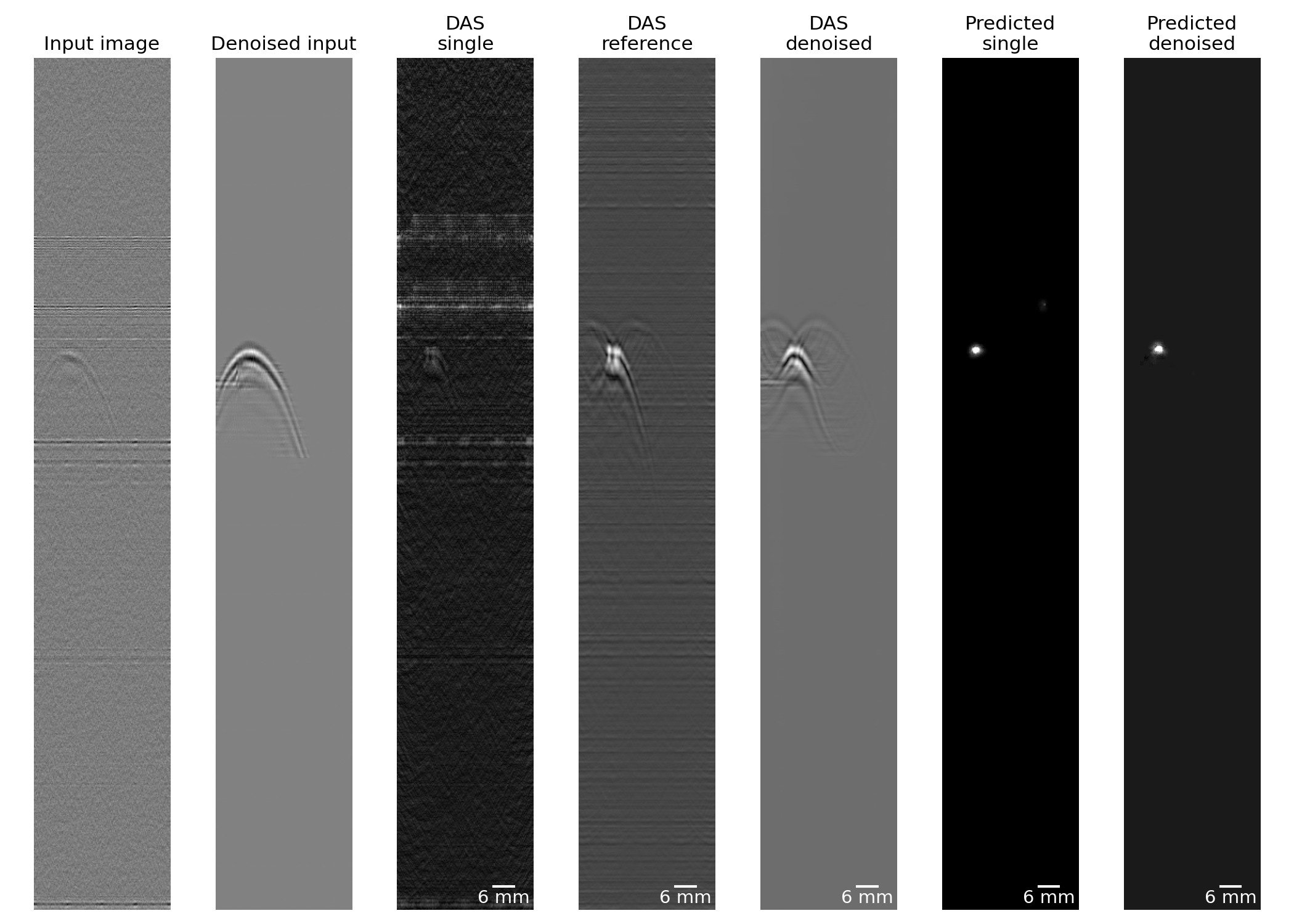}\hfill
    \caption{}
    \end{subfigure}
    \begin{subfigure}{\linewidth}
    \includegraphics[width=\linewidth]{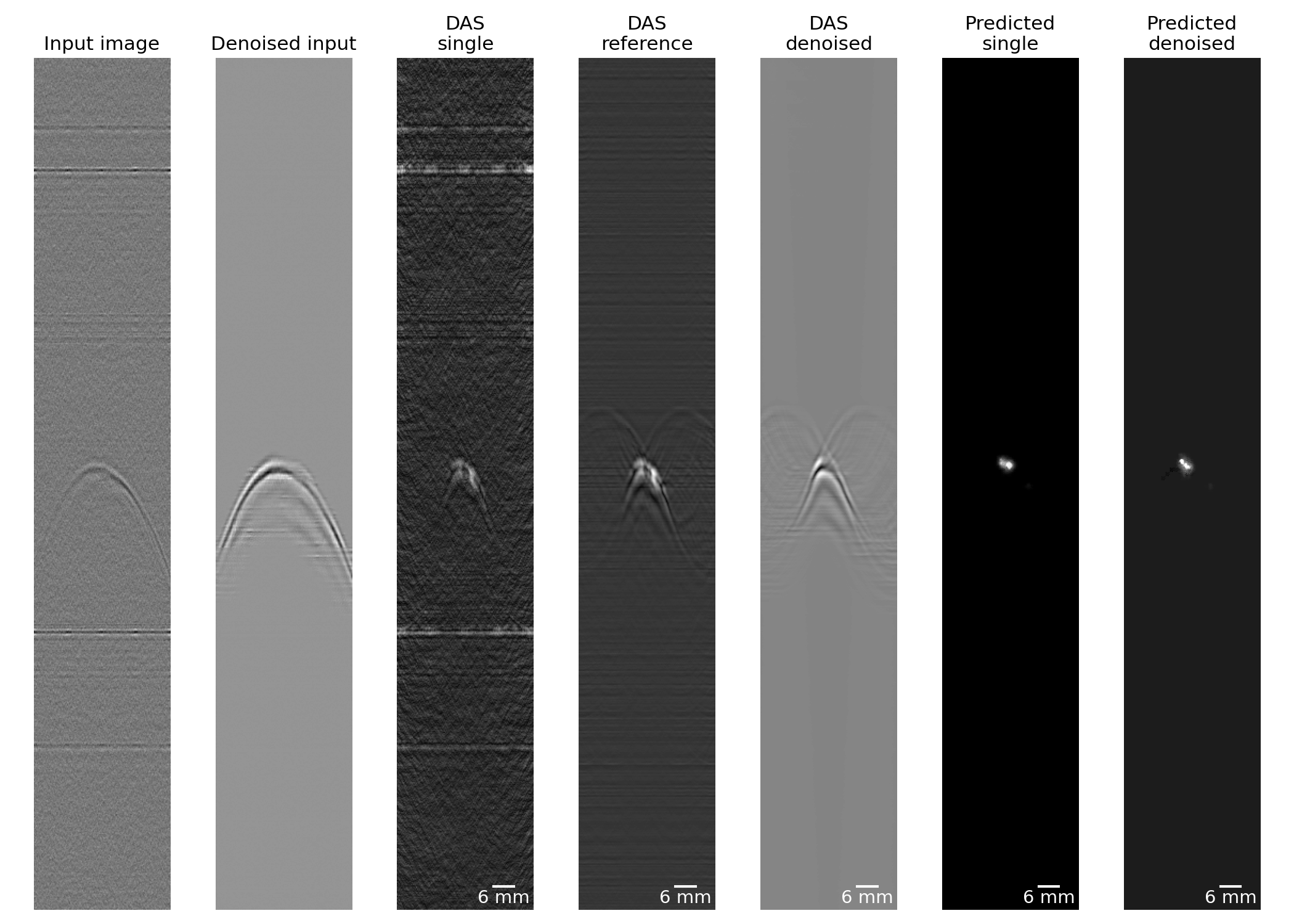}\hfill
    \caption{}
    \end{subfigure}
    \caption{Successful predictions of the reconstruction model from a single frame.}
    \label{fig:reconstr-1}
\end{figure}

\begin{figure}[!t]
    \centering
    \begin{subfigure}{\linewidth}
    \includegraphics[width=\linewidth]{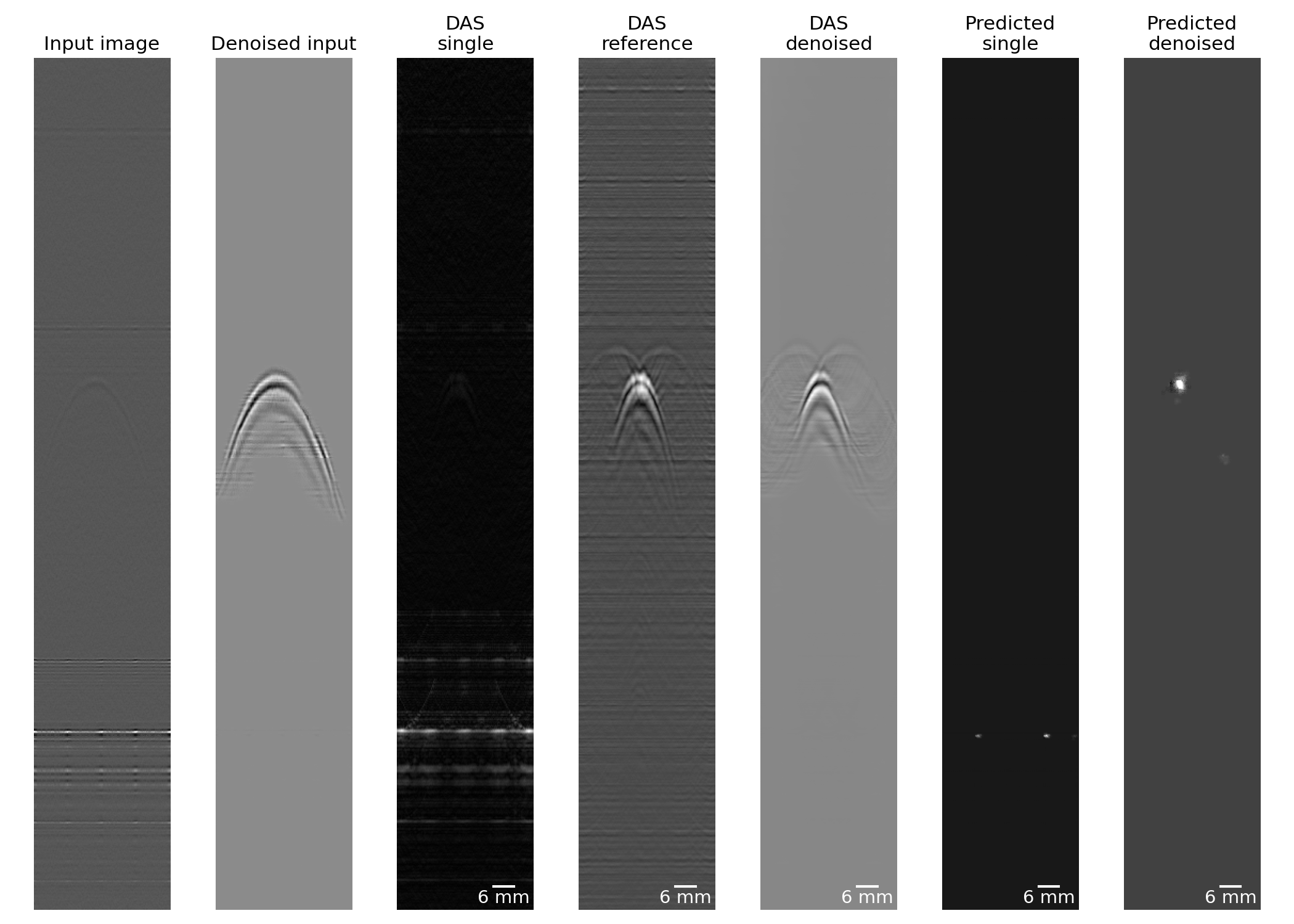}\hfill
    \caption{}
    \end{subfigure}
    \begin{subfigure}{\linewidth}
    \includegraphics[width=\linewidth]{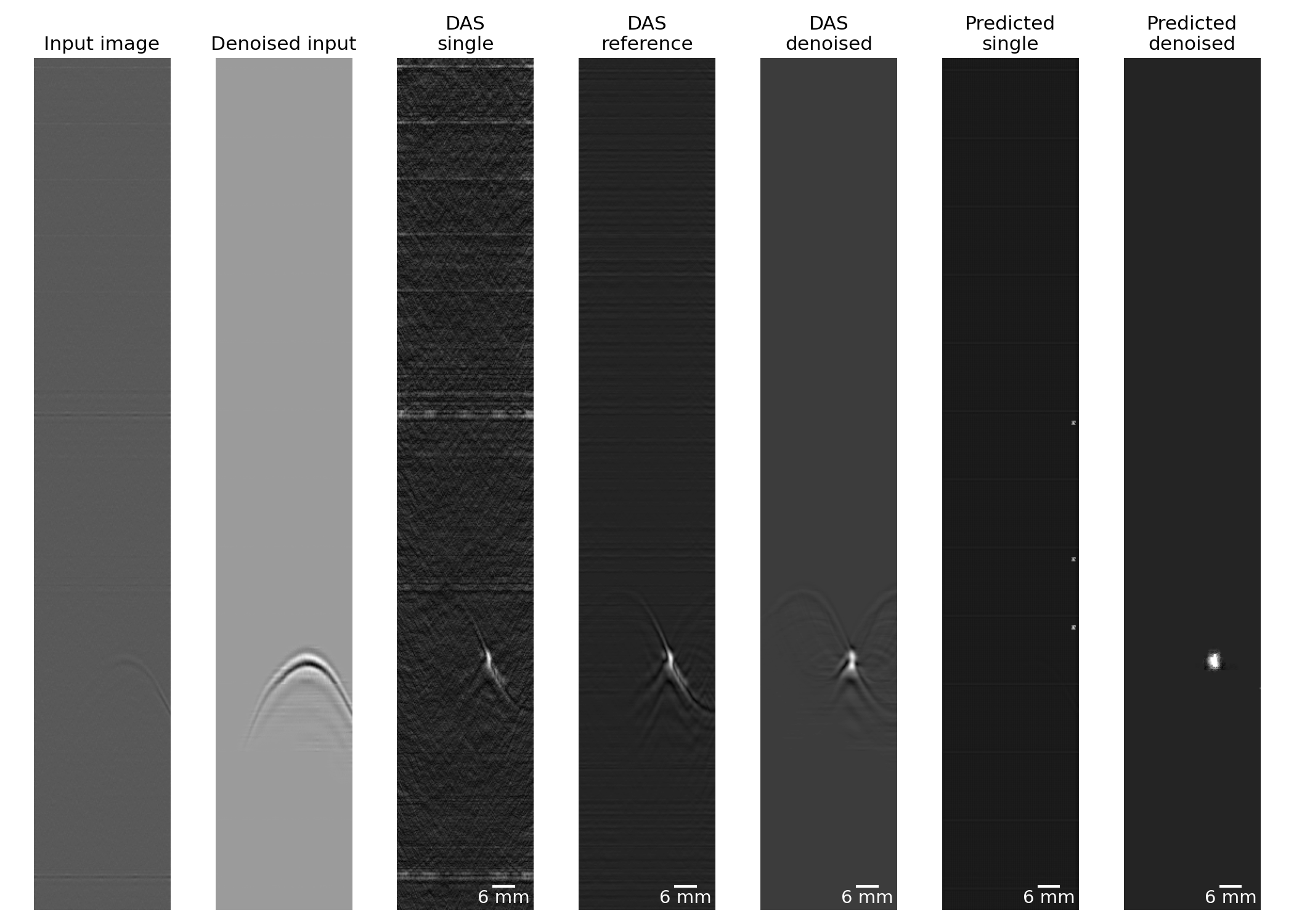}\hfill
    \caption{}
    \end{subfigure}
    \caption{Successful predictions from Dual-GAN model from a single frame.}
    \label{fig:reconstr-2}
\end{figure}

In Figures \ref{fig:reconstr-1} and \ref{fig:reconstr-2}, results of the reconstruction model are shown. Four samples were randomly chosen, where two samples showed a successful reconstruction and two samples showed successful Dual-GAN reconstruction. In the samples shown in Figure \ref{fig:reconstr-1}, a point source was predicted from a single frame. In the samples in Figure \ref{fig:reconstr-2}, no point source was predicted from a single frame, but a point source was predicted using Dual-GAN. From 20 samples in the test set, 14 samples had a successful prediction using only the reconstruction model. For the remaining six samples, Dual-GAN managed to predict the reconstruction correctly.

For each image set, there are 7 columns, showing from left to right: the single frame that was used as input to the model, the prediction of the Pix2Pix-Residual denoising model, trained on segmented data, the DAS reconstruction calculated from a single frame, the DAS reconstruction from the reference image, the DAS reconstruction calculated from the denoised image, the predicted reconstruction by the Pix2Pix reconstruction model, and the reconstruction based on the Dual-GAN model. The Pix2Pix-Residual model trained on segmented data was used for the Dual-GAN approach here, based on the sharp parabola shape and little artifacts, which is essential for a good reconstruction.

\section{Discussion}\label{sec:discussion}
Table \ref{tab:metrics-denoise} shows the metrics of the different denoising methods. Here, it is visible that every model, with the exception of Pix2Pix-Residual with reference images as target data, improved in terms of SSIM compared to the noisy data. Furthermore, both Pix2Pix models improved in terms of MSE, with the Pix2Pix-Residual trained on segmented data falling closely behind. Looking at the time performance, we found that the Pix2Pix models can achieve frame rates between 25-32 Hz on GPU and 20-22 Hz on CPU, whereas the Pix2Pix-Residual models fall clearly behind with 13-16 Hz on GPU and around 1-2 Hz on CPU. Based on these metrics, it appears that the Pix2Pix models generally outperform the Pix2Pix-Residual models, as well as averaging techniques. However, as noted before, the metrics for the models trained on segmented images were evaluated with the reference images as ground truth, which may cause the metrics for these models to be worse, while performance is better. For this reason, we also carried out a visual evaluation.

\afterpage{\FloatBarrier}  

In Figure \ref{fig:denoise-p2p}, three predictions, randomly chosen from the test data, using the Pix2Pix model trained on reference images are shown, along with the corresponding input and reference images. Here, it is apparent that the signal strength has increased compared to the input image. Interestingly, it appears that the model has also learned to slightly improve the signal strength compared to the reference image. However, we can see that the background is a bit distorted, especially around the characteristic parabola. We have not yet identified the cause of this problem, however, what may have caused this is that the background of the reference patches are not completely empty; since our dataset is purely based on experimental data, the only reference that we have is based on averaging many frames. However, this does not perfectly remove all noise, and still leaves some artifacts in the `background' patches as well. 

To tackle this problem, and given the viability of our method with little available data, the Pix2Pix model was retrained with segmented images as target data. In other words, the model was trained to learn background patches to become empty, instead of the average that was previously used. The predictions of this newly trained model on the test set are shown in Figure \ref{fig:denoise-p2p-noBG}. Here, it is visible that the background is much better predicted in the image. The contrast of the signal parabola is similar to the contrast seen in Figure \ref{fig:denoise-p2p}, but here, the background looks much more homogeneous. 

The results for the Pix2Pix-Residual model trained with reference images as targets are shown in Figure \ref{fig:denoise-p2pr}. Here, the model is still able to improve the signal strength, but the background looks much more distorted compared to the results in Figure \ref{fig:denoise-p2p}. However, the characteristic parabolas look sharper and with less artifacts, which were visible in Figure \ref{fig:denoise-p2p}a and \ref{fig:denoise-p2p}c. To tackle the background problem, the Pix2Pix-Residual model was trained with segmented images as target data too. The results of this model are shown in Figure \ref{fig:denoise-p2pr-noBG}. Here, we see that the characteristic parabola is predicted very well, with no artifacts and a better prediction of the secondary waves, which were predicted to be more blurry by the Pix2Pix model.

Looking at all denoising results, it is clear that all models (except Pix2Pix-Residual with reference images as target) managed to improve the input image in terms of contrast and signal strength. The Pix2Pix model generally outperformed the Pix2Pix-Residual models in metrics and time performance, however, the Pix2Pix-Residual model shows good performance when evaluating the images visibly, especially looking at the secondary waves in the image and the clearness of the parabola. However, the quality of the input image is an important part of the predictions. When the input signal is too weak, no prediction or half a prediction is made. A good example of a `half prediction' is visible in Figure \ref{fig:denoise-p2p}c, where the input image does not show a clear parabola line on the left side, and in return, the model is unable to predict this half of the parabola.

In Figure \ref{fig:reconstr-1}, some successful predictions by the reconstruction model are shown. It is visible that the direct reconstructions do well here and able to estimate a point at the centre location of the DAS reconstruction. Dual-GAN shows a similar effect here. 
However, in 6 out of 20 test samples, no point source could be predicted from a single frame of RF data. In these cases, we applied the Dual-GAN method, in which the Pix2Pix-Residual model was employed to denoise the input first, then, use this denoised input to predict the point source. In all six failure cases, this method allowed for accurate prediction for the point source. Two examples are shown in Figure \ref{fig:reconstr-2}. In Figure \ref{fig:reconstr-2}a, it is also visible that there is a low-intensity point source predicted to the right bottom of the actual point source, whereas there is not supposed to be a point source, as apparent from the DAS reconstructions. This artifact can be possibly attributed to the accumulating error principle; when the denoising model predicts the parabola at a different position, this error is propagated into the reconstruction model. The total error becomes the sum of both denoising and reconstruction models. Another possible drawback of Dual-GAN, and possible cause of this artifact, is that the reconstruction model was trained with low-SNR inputs, whereas the denoising model gives high-SNR outputs by design. Usually, the reconstruction model does handle this well, however, it is possible that the artifact from Figure \ref{fig:reconstr-2}a was caused by this mismatch in input data.

\section{Conclusion}\label{sec:conclusion}
In this study, we aimed to utilize DL-based models for pre-beamformed denoising and reconstruction of PA point source images. To this end, in addition to the conventional averaging method, two DL-based models were employed and compared: the Pix2Pix model, and an adaptation to this model, named Pix2Pix-Residual. These models were trained on two target datasets, to compare performance with different background patterns. Performance for denoising using Pix2Pix showed very promising results, as well as the Pix2Pix-Residual model when trained on target data where background noise was manually removed. Automated reconstruction using Pix2Pix has also proven possible, although in some cases, the signal intensity was too low to predict the location of the point source. In these situations, a point source can be predicted by combining the reconstruction algorithm with the denoising algorithm sequentially.

\bibliography{main}

\begin{thebibliography}{10}
\newcommand{\enquote}[1]{``#1''}

\bibitem{Attia2019}
A.~B.~E. Attia, G.~Balasundaram, M.~Moothanchery, U.~S. Dinish, R.~Bi,
  V.~Ntziachristos, and M.~Olivo, \enquote{{A review of clinical photoacoustic
  imaging: Current and future trends},} {\protect\JournalTitle{Photoacoustics}}
  \textbf{16}, 100144 (2019).

\bibitem{beard2011biomedical}
P.~Beard, \enquote{Biomedical photoacoustic imaging,}
  {\protect\JournalTitle{Interface focus}} \textbf{1}, 602--631 (2011).

\bibitem{Jeon2019a}
S.~Jeon, J.~Kim, D.~Lee, J.~W. Baik, and C.~Kim, \enquote{{Review on practical
  photoacoustic microscopy},}  (2019).

\bibitem{Steinberg2019}
I.~Steinberg, D.~M. Huland, O.~Vermesh, H.~E. Frostig, W.~S. Tummers, and S.~S.
  Gambhir, \enquote{{Photoacoustic clinical imaging},}  (2019).

\bibitem{Xu2006}
M.~Xu and L.~V. Wang, \enquote{{Photoacoustic imaging in biomedicine},}
  {\protect\JournalTitle{Review of Scientific Instruments}} \textbf{77}, 041101
  (2006).

\bibitem{Xu2006PhotoacousticBiomedicine}
M.~Xu and L.~V. Wang, \enquote{{Photoacoustic imaging in biomedicine},}
  {\protect\JournalTitle{Review of Scientific Instruments}} \textbf{77} (2006).

\bibitem{Yang2011}
Y.~Yang, X.~Li, T.~Wang, P.~D. Kumavor, A.~Aguirre, K.~K. Shung, Q.~Zhou,
  M.~Sanders, M.~Brewer, and Q.~Zhu, \enquote{{Integrated optical coherence
  tomography, ultrasound and photoacoustic imaging for ovarian tissue
  characterization},} {\protect\JournalTitle{Biomedical Optics Express}}
  \textbf{2}, 2551 (2011).

\bibitem{Gao2014}
F.~Gao, X.~Feng, Y.~Zheng, and C.-D. Ohl, \enquote{{Photoacoustic resonance
  spectroscopy for biological tissue characterization},}
  {\protect\JournalTitle{Journal of Biomedical Optics}} \textbf{19}, 067006
  (2014).

\bibitem{Su2010PhotoacousticTissue}
J.~Su, A.~Karpiouk, B.~Wang, and S.~Emelianov, \enquote{{Photoacoustic imaging
  of clinical metal needles in tissue},} {\protect\JournalTitle{Journal of
  Biomedical Optics}} \textbf{15}, 021309 (2010).

\bibitem{LedijuBell2018Photoacoustic-basedTip}
M.~A. Lediju~Bell and J.~Shubert, \enquote{{Photoacoustic-based visual servoing
  of a needle tip},} {\protect\JournalTitle{Scientific Reports}} \textbf{8},
  1--12 (2018).

\bibitem{Dogra2013MultispectralResults}
V.~Dogra, B.~Chinni, K.~Valluru, J.~Joseph, A.~Ghazi, J.~Yao, K.~Evans,
  E.~Messing, and N.~Rao, \enquote{{Multispectral photoacoustic imaging of
  prostate cancer: Preliminary ex-vivo results},}
  {\protect\JournalTitle{Journal of Clinical Imaging Science}} \textbf{3}
  (2013).

\bibitem{Ishihara2017}
M.~Ishihara, M.~Shinchi, A.~Horiguchi, H.~Shinmoto, H.~Tsuda, K.~Irisawa,
  T.~Wada, and T.~Asano, \enquote{{Possibility of transrectal photoacoustic
  imaging-guided biopsy for detection of prostate cancer},} in \emph{Photons
  Plus Ultrasound: Imaging and Sensing 2017,}  vol. 10064 (SPIE, 2017), p.
  100642U.

\bibitem{LedijuBell2015TransurethralImaging}
M.~A. Lediju~Bell, X.~Guo, D.~Y. Song, and E.~M. Boctor,
  \enquote{{Transurethral light delivery for prostate photoacoustic imaging},}
  {\protect\JournalTitle{Journal of Biomedical Optics}} \textbf{20}, 036002
  (2015).

\bibitem{Kolkman2006InDiode}
R.~G. Kolkman, W.~Steenbergen, and T.~G. Van~Leeuwen, \enquote{{In vivo
  photoacoustic imaging of blood vessels with a pulsed laser diode},}
  {\protect\JournalTitle{Lasers in Medical Science}} \textbf{21}, 134--139
  (2006).

\bibitem{Upputuri2018FastReview}
P.~K. Upputuri and M.~Pramanik, \enquote{{Fast photoacoustic imaging systems
  using pulsed laser diodes: a review},}  (2018).

\bibitem{Moradi2021ARecovery}
H.~Moradi, Y.~Wu, S.~E. Salcudean, and E.~Boctor, \enquote{{A photoacoustic
  image reconstruction method for point source recovery},} in \emph{Photons
  Plus Ultrasound: Imaging and Sensing 2021,}  vol. 11642 (SPIE, 2021), p.~61.

\bibitem{Wang2019}
H.~Wang, S.~Liu, T.~Wang, C.~Zhang, T.~Feng, and C.~Tian,
  \enquote{{Three-dimensional interventional photoacoustic imaging for biopsy
  needle guidance with a linear array transducer},}
  {\protect\JournalTitle{Journal of Biophotonics}} \textbf{12}, e201900212
  (2019).

\bibitem{Kim2018MultimodalGuidance}
H.~Kim and J.~H. Chang, \enquote{{Multimodal photoacoustic imaging as a tool
  for sentinel lymph node identification and biopsy guidance},}  (2018).

\bibitem{Moradi2020Robot-assistedSurgery}
H.~Moradi, E.~M. Boctor, and S.~E. Salcudean, \enquote{{Robot-assisted image
  guidance for prostate nerve-sparing surgery},} {\protect\JournalTitle{IEEE
  International Ultrasonics Symposium, IUS}} \textbf{2020-September} (2020).

\bibitem{Song2022Real-timeDemonstration}
H.~Song, H.~Moradi, B.~Jiang, K.~Xu, Y.~Wu, R.~H. Taylor, A.~Deguet, J.~U.
  Kang, S.~E. Salcudean, and E.~M. Boctor, \enquote{{Real-time intraoperative
  surgical guidance system in the da Vinci surgical robot based on transrectal
  ultrasound/photoacoustic imaging with photoacoustic markers: an ex vivo
  demonstration},} {\protect\JournalTitle{IEEE Robotics and Automation
  Letters}}  (2022).

\bibitem{Roehl2004}
K.~A. Roehl, M.~Han, C.~G. Ramos, J.~A.~V. Antenor, and W.~J. Catalona,
  \enquote{{Cancer progression and survival rates following anatomical radical
  retropubic prostatectomy in 3,478 consecutive patients: Long-term results},}
  {\protect\JournalTitle{Journal of Urology}} \textbf{172}, 910--914 (2004).

\bibitem{Badani2007}
K.~K. Badani, S.~Kaul, and M.~Menon, \enquote{{Evolution of robotic radical
  prostatectomy: Assessment after 2766 procedures},}
  {\protect\JournalTitle{Cancer}} \textbf{110}, 1951--1958 (2007).

\bibitem{Refaee2021DenoisingNetworks}
A.~Refaee, C.~J. Kelly, H.~Moradi, and S.~E. Salcudean, \enquote{{Denoising of
  pre-beamformed photoacoustic data using generative adversarial networks},}
  {\protect\JournalTitle{Biomedical Optics Express}} \textbf{12}, 6184 (2021).

\bibitem{pix2pix2017}
P.~Isola, J.-Y. Zhu, T.~Zhou, and A.~A. Efros, \enquote{Image-to-image
  translation with conditional adversarial networks,}
  {\protect\JournalTitle{CVPR}}  (2017).

\bibitem{Kim2020Deep-LearningSystem}
M.~W. Kim, G.~S. Jeng, I.~Pelivanov, and M.~O'Donnell, \enquote{{Deep-Learning
  Image Reconstruction for Real-Time Photoacoustic System},}
  {\protect\JournalTitle{IEEE Transactions on Medical Imaging}} \textbf{39},
  3379--3390 (2020).

\bibitem{Hauptmann2020DeepDirections}
A.~Hauptmann and B.~T. Cox, \enquote{{Deep learning in photoacoustic
  tomography: current approaches and future directions},}
  {\protect\JournalTitle{https://doi.org/10.1117/1.JBO.25.11.112903}}
  \textbf{25}, 112903 (2020).

\bibitem{Lan2019Ki-GAN:Vivo}
H.~Lan, K.~Zhou, C.~Yang, J.~Cheng, J.~Liu, S.~Gao, and F.~Gao,
  \enquote{{Ki-GAN: Knowledge Infusion Generative Adversarial Network for
  Photoacoustic Image Reconstruction In Vivo},} in \emph{Lecture Notes in
  Computer Science,}  vol. 11764 LNCS (Springer Science and Business Media
  Deutschland GmbH, 2019), pp. 273--281.

\bibitem{Lan2020Y-Net:Vivo}
H.~Lan, D.~Jiang, C.~Yang, F.~Gao, and F.~Gao, \enquote{{Y-Net: Hybrid deep
  learning image reconstruction for photoacoustic tomography in vivo},}
  {\protect\JournalTitle{Photoacoustics}} \textbf{20}, 100197 (2020).

\bibitem{Guo2022AS-Net:Data}
M.~Guo, H.~Lan, C.~Yang, J.~Liu, and F.~Gao, \enquote{{AS-Net: Fast
  Photoacoustic Reconstruction with Multi-Feature Fusion from Sparse Data},}
  {\protect\JournalTitle{IEEE Transactions on Computational Imaging}}
  \textbf{8}, 215--223 (2022).

\bibitem{Waibel2018ReconstructionLearning}
D.~Waibel, J.~Gr{\"{o}}hl, F.~Isensee, T.~Kirchner, K.~Maier-Hein, and
  L.~Maier-Hein, \enquote{{Reconstruction of initial pressure from limited view
  photoacoustic images using deep learning},} in \emph{Photons Plus Ultrasound:
  Imaging and Sensing 2018,}  vol. 10494 A.~A. Oraevsky and L.~V. Wang, eds.
  (SPIE, 2018), p.~98.

\bibitem{Reiter2017AData}
A.~Reiter and M.~A. Lediju~Bell, \enquote{{A machine learning approach to
  identifying point source locations in photoacoustic data},} in \emph{Photons
  Plus Ultrasound: Imaging and Sensing 2017,}  vol. 10064 (SPIE, 2017), p.
  100643J.

\bibitem{Johnstonbaugh2020AMedium}
K.~Johnstonbaugh, S.~Agrawal, D.~A. Durairaj, C.~Fadden, A.~Dangi, S.~P.~K.
  Karri, and S.~R. Kothapalli, \enquote{{A Deep Learning Approach to
  Photoacoustic Wavefront Localization in Deep-Tissue Medium},}
  {\protect\JournalTitle{IEEE Transactions on Ultrasonics, Ferroelectrics, and
  Frequency Control}} \textbf{67}, 2649--2659 (2020).

\bibitem{Johnstonbaugh2019NovelLocalization}
K.~Johnstonbaugh, S.~Agrawal, D.~A. Durairaj, M.~Homewood, S.~P. Krisna~Karri,
  and S.-R. Kothapalli, \enquote{{Novel deep learning architecture for optical
  fluence dependent photoacoustic target localization},} in \emph{Photons Plus
  Ultrasound: Imaging and Sensing 2019,}  vol. 10878 (SPIE, 2019), p.~55.

\bibitem{Allman2018PhotoacousticLearning}
D.~Allman, A.~Reiter, and M.~A. Bell, \enquote{{Photoacoustic Source Detection
  and Reflection Artifact Removal Enabled by Deep Learning},}
  {\protect\JournalTitle{IEEE Transactions on Medical Imaging}} \textbf{37},
  1464--1477 (2018).

\bibitem{Jeon2019}
S.~Jeon, E.~Y. Park, W.~Choi, R.~Managuli, K.~j. Lee, and C.~Kim,
  \enquote{{Real-time delay-multiply-and-sum beamforming with coherence factor
  for in vivo clinical photoacoustic imaging of humans},}
  {\protect\JournalTitle{Photoacoustics}} \textbf{15}, 100136 (2019).

\bibitem{vanBoxtel2021HybridImages}
J.~van Boxtel, V.~Vousten, J.~Pluim, and N.~M. Rad, \enquote{{Hybrid Deep
  Neural Network for Brachial Plexus Nerve Segmentation in Ultrasound Images},}
  in \emph{2021 29th European Signal Processing Conference (EUSIPCO),}  (2021),
  pp. 1246--1250.

\bibitem{Xu2017Dual-stageVasculopathy}
Y.~Xu, K.~Yan, J.~Kim, X.~Wang, C.~Li, L.~Su, S.~Yu, X.~Xu, and D.~D. Feng,
  \enquote{Dual-stage deep learning framework for pigment epithelium detachment
  segmentation in polypoidal choroidal vasculopathy,}
  {\protect\JournalTitle{Biomed. Opt. Express}} \textbf{8}, 4061--4076 (2017).

\bibitem{Ying2019AnSolutions}
X.~Ying, \enquote{{An Overview of Overfitting and its Solutions},}
  {\protect\JournalTitle{Journal of Physics: Conference Series}} \textbf{1168},
  022022 (2019).

\bibitem{Ronneberger2015U-NetSegmentation}
O.~Ronneberger, P.~Fischer, and T.~Brox, \enquote{{U-Net: Convolutional
  Networks for Biomedical Image Segmentation},} {\protect\JournalTitle{CoRR}}
  \textbf{abs/1505.0}, 234--241 (2015).

\bibitem{Chi2020denoising}
J.~Chi, C.~Wu, X.~Yu, P.~Ji, and H.~Chu, \enquote{Single low-dose ct image
  denoising using a generative adversarial network with modified u-net
  generator and multi-level discriminator,} {\protect\JournalTitle{IEEE
  Access}} \textbf{8}, 133470--133487 (2020).

\bibitem{Han2018FramingCT}
Y.~Han and J.~C. Ye, \enquote{{Framing U-Net via Deep Convolutional Framelets:
  Application to Sparse-View CT},} {\protect\JournalTitle{IEEE Transactions on
  Medical Imaging}} \textbf{37}, 1418--1429 (2018).

\bibitem{Isola2017Image-to-imageNetworks}
P.~Isola, J.~Y. Zhu, T.~Zhou, and A.~A. Efros, \enquote{{Image-to-image
  translation with conditional adversarial networks},} in \emph{Proceedings -
  30th IEEE Conference on Computer Vision and Pattern Recognition, CVPR 2017,}
  vol. 2017-Janua (2017), pp. 5967--5976.

\bibitem{Gurrola-Ramos2021ADenoising}
J.~Gurrola-Ramos, O.~Dalmau, and T.~E. Alarcon, \enquote{{A Residual Dense
  U-Net Neural Network for Image Denoising},} {\protect\JournalTitle{IEEE
  Access}} \textbf{9}, 31742--31754 (2021).

\bibitem{Wang2004ImageSimilarity}
Z.~Wang, A.~C. Bovik, H.~R. Sheikh, and E.~P. Simoncelli, \enquote{{Image
  Quality Assessment: From Error Visibility to Structural Similarity},}
  {\protect\JournalTitle{IEEE Transactions on Image Processing}} \textbf{13},
  600--612 (2004).

\end{thebibliography}

\end{document}